\documentclass[useAMS,usenatbib]{mn2e}
\pdfoutput=1
\usepackage{journals}
\usepackage[pdftex]{graphicx,color}
\usepackage[latin1]{inputenc}
\usepackage{graphics}
\usepackage{amsfonts}
\usepackage{amsmath}
\usepackage{multicol}
\usepackage{layout}
\usepackage{amssymb}
\usepackage[a4paper,colorlinks=true,pdfstartview=FitV,
linkcolor=red,citecolor=blue,urlcolor=magenta,draft]{hyperref}
 \title[Halo Shapes Evolution]
{Some like  it Triaxial: the  universality of Dark Matter  Halo shapes
  and their  evolution along the  cosmic time} \author[Despali  et al.
2014]      {\parbox{\textwidth}{Giulia      Despali$^1$\thanks{E-mail:
      \href{mailto:giulia.despali@studenti.unipd.it}
      {giulia.despali@studenti.unpd.it}}, Carlo Giocoli$^{2,3,4}$,
    Giuseppe Tormen$^1$}\\
  $^1$ Dipartimento di Fisica e Astronomia, Universit\`{a} degli Studi
  di Padova, vicolo dell'Osservatorio 3, 35122,
  Padova, Italy\\
  $^2$  Dipartimento  di Fisica  e  Astronomia,  Alma Mater  Studiorum
  Universit\`{a} di Bologna, viale Berti Pichat, 6/2, 40127
  Bologna, Italy\\
  $^3$ INAF - Osservatorio Astronomico di
  Bologna, via Ranzani 1, 40127, Bologna, Italy \\
  $^4$  INFN -  Sezione di  Bologna,  viale Berti  Pichat 6/2,  40127,
  Bologna, Italy }
       
\begin{document}
\date{}
\maketitle
\label{firstpage}
\pagerange{\pageref{firstpage}--\pageref{lastpage}} \pubyear{2010}
\begin{abstract}
  We present a detailed analysis  of dark matter halo shapes, studying
  how the  distributions of ellipticity, prolateness  and axial ratios
  evolve as a  function of time and mass.  With  this purpose in mind,
  we analysed  the results of three  cosmological simulations, running
  an ellipsoidal  halo finder  to measure  triaxial halo  shapes.  The
  simulations  have different  scales,  mass  limits and  cosmological
  parameters,  which  allows  us  to  ensure  a  good  resolution  and
  statistics in a  wide mass range, and to  investigate the dependence
  of halo properties on the cosmological model.

  We confirm the  tendency of haloes to be prolate  at all times, even
  if they  become more triaxial  going to higher  redshifts. Regarding
  the dependence on mass, more  massive haloes are also less spherical
  at all redshifts, since they are the most recent forming systems and
  so  still retain memory  of their  original shape  at the  moment of
  collapse. We  then propose a rescaling of  the shape-mass relations,
  using  the variable $\nu=\delta_{c}/\sigma$  to represent  the mass,
  which absorbs the dependence on both cosmology and time, allowing to
  find universal  relations between  halo masses and  shape parameters
  (ellipticity, prolateness  and the axial  ratios) which hold  at any
  redshift. This  may be very useful to  determine prior distributions
  of halo shapes for observational studies.

\end{abstract}

\section{introduction}
Nowadays  different  observational  campaigns agree  on  the  standard
cosmological model to explain and  describe the structure formation in
our Universe  \citep{fu08,planckxx}.  In this scenario,  almost $95\%$
of the  energy content on the  Universe is in unknown  forms of energy
and  matter,  generally  called  dark energy  and  dark  matter.   The
structure formation  process occurs  around the initial  density peaks
\citep{bardeen86,bond96,ludlow11,paranjape12,paranjape13} and proceeds
hierarchically along the cosmic time: small systems collapse first, at
high  redshift, and  then  merge together  forming  more massive  ones
\citep{lacey93,lacey94,tormen98a}.   Galaxy clusters  are the  largest
virialized systems  in the Universe  and so  the last to  form: almost
$80\%$ of their  mass is attributed to dark matter,  while the rest to
baryons, divided in $hot$ (diffuse gas, $75\%$), $cold$ (stars, $7\%$)
and  other   forms  ($18\%$)   \citep{ettori09}.   They   collapse  as
consequence  of gravitational  instability  and grow  as  a result  of
different violent  merging events \citep{tormen04};  many observations
have  captured  them  as  characterised by  multiple  mass  components
\citep{postman12,zitrin13a},  elongated  in  the   plane  of  the  sky
\citep{zitrin13b}  or   along  the   line-of-sight  \citep{morandi11}.
Galaxy  clusters  are  also  characterised by  the  presence  of  many
substructures,  which  are the  cores  of  progenitor haloes  accreted
during the  formation history and  may contain galaxy  cluster members
\citep{giocoli10}.  The  mass density  distribution of  relaxed haloes
typically follows a well defined  profile \citep{navarro96} that has a
logarithmic  slope of  $-1$  in the  inner part  and  $-3$ toward  the
outskirt.  The distance from the center at which the logarithmic slope
approaches  $-2$ defines  the scale  radius  $r_s$ from  which we  can
define the  concentration $c=R_{vir}/r_s$, where  $R_{vir}$ represents
the virialisation radius  of the system.  Galaxy clusters  can be used
as  cosmological probes  since  they  are expected  to  follow a  well
defined                   concentration-mass                  relation
\citep{neto07,zhao09,prada11,giocoli12b} and their predicted abundance
as     a     function     of    redshift     is     well     portrayed
\citep{sheth99b,jenkins01,tinker08}.  These may  then be combined with
other       analyses       for        example       cosmic       shear
\citep{schrabback07,schrabback10}, CMB  primordial density fluctuation
\citep{planckxx},   Sunyaev   Zel'dovich   effect   and   X-ray   data
\citep{roncarelli07,roncarelli10}, two \citep{marulli12,marulli13} and
three point correlation functions \citep{moresco13}.

However, the use of galaxy  clusters as cosmological probes depends on
how well structural properties -- mass, concentration, triaxiality and
subhalo   abundance   --   can   be  recovered   combining   different
observations.   Many studies  have  revealed that  the possibility  to
correctly estimate the  mass and the concentration of  the clusters is
an  open problem  for X-ray,  SZ and  also optical  observations.  For
example, different lensing analyses on numerical and pseudo-analytical
clusters have shown that the estimated mass is on average biased to be
lower than  the true one \citep{becker11,meneghetti10b}.   This is due
to the  fact that  the dark  matter haloes in  which the  clusters are
embedded are  typically prolate \citep{giocoli12c}, while  most of the
works still uses spherical models:  we have a large probability to see
them elongated in the plane of the sky and this leads to underestimate
the mass if we limit ourselves to spherical halo boundaries which will
not capture the whole halo.  On the other hand, when the major axis of
the halo  ellipsoid is  elongated along the  line-of-sight we  tend to
overestimate the mass.  A correct  modeling of the halo triaxiality is
therefore important in order to  reach a better understanding of these
biases  and  how  structural  properties  can be  recovered  from  the
observations.  The triaxiality of the halo is also responsible for the
amplification of  the Einstein radius  size (both when the  cluster is
very elongated  in the plane of  the sky and  along the line-of-sight)
and of the generation a  possibile bias in the estimated concentration
and  in the  measurement of  the inner  slope of  the  density profile
\citep{giocoli14}.  SZ and X-ray  mass reconstructions tend to rely on
the assumption that the systems  are spherically symmetric and the hot
gas is  in hydrostatic  equilibrium.  \citet{rasia12} have  shown that
the  X-ray  mass  are  on  average  biased  low  by  a  large  amount,
highlighting both the presence of non-thermal pressure and temperature
anisotropy in the  inter cluster medium.  In this  context it is worth
to  notice that  the discrepancy  between X-ray,  SZ and  lensing mass
appear not only when the cluster is not relaxed but also when its mass
density  distribution  deviate   from  spherical  symmetry.   Combined
multi-wavelength   observations  are  thus   important  not   only  to
understand  the physical  state of  cosmic structures  but  also their
shape.

The  interpretation of  observations  requires a  comparison with  the
predictions  coming from  theory  and simulations  \citep{limousin13}.
For this reason,  it is becoming more and more  important to model the
results of simulations  with as much detail as  possible, even if this
is  computationally more  expensive.  Dark  matter haloes  are usually
identified  in simulations  as spherical  systems, since  it  is quite
easy, computationally  speaking, and  it is also  proven to be  a good
approximation  when  calculating  the  main  properties  of  the  halo
population,  such  as the  mass  function,  the  concentration and  so
on. Moreover, even if their boundaries are spherical, obviously matter
inside  them is  not  isotropically  distributed and  so  even from  a
spherical distribution it is possible  to compute the axial ratios and
other shape  parameters.  Nevertheless,  it is clear  that considering
all systems as spheres is a bit rough and so many works claim the need
of   a   more   realistic   model,   such   as   triaxial   ellipsoids
\citep{warren92,jing02,allgood06,despali13}.  Being  more precise  and
allowing  a greater  variety of  shapes, this  method  is particularly
useful when one  wants to study dark matter  halo shapes and determine
their influence on observable 2D projected quantities.

A  precise  knowledge  of  the  ellipticity and  of  the  axial  ratio
distributions of  galaxy cluster-size  halos is fundamental  also when
observations from different bands are  combined to recover the cluster
mass  \citep{morandi10,morandi12}.   In  this  case  the  distribution
priors can be  used to constrain both the ellipticity  on the plane of
the  sky and  the elongation  along the  line-of-sight of  the cluster
ellipsoid \citep{sereno13}.

Thus, the  aim of this work is  to study the shape  of triaxial haloes
and  model  its distribution,  its  evolution  with  redshift and  the
dependence  on   cosmology.   In  particular,  we   will  analyse  the
distribution of  the shape  parameters (axial ratios,  ellipticity and
prolateness) as  a function  of halo mass  and redshift; we  will then
present  some universal relations  and fitting  formulae which  may be
used to retrieve  the typical shape distribution at  a certain time or
for  a certain  mass bin,  when a  comparison with  observations  or a
prediction is needed.

The   structure   of  this   paper   is   as  follows.    In   Section
\ref{simulations} we  present the  numerical simulations used  in this
work  and  the  post   processing  pipeline.   Section  \ref{results1}
describes the results  on shape parameters: we  analyse the properties
of the  halo population  at various redshifts  of the  simulations and
show  some universal  fitting formulae  for the  shape parameters.  In
Section \ref{mergerzf} we describe the  merger tree catalogues and how
it is  possible to  relate the  shape with  the formation  redshift of
haloes.  Finally in Section \ref{conclusions} we summarise and discuss
our results.

\section{The Numerical Simulations}\label{simulations}

\begin{table*}
\centering
\begin{tabular}{|c|c|c|c|c|c|c|c|c|c|c|}
  \hline
  &$\Omega_{m}$ &$\Omega_{\Lambda}$ & H[km $s^{-1}$] & box [Mpc $h^{-1}$]
  & $z_{i}$ & N & $m_{p}$[$M_{\odot}h^{-1}$] & soft [kpc $h^{-1}$] &$\sigma_{8}$ &
  $M_{*}(0)$[$10^{12}M_{\odot}h^{-1}$]\\
  \hline
  \textbf{GIF2} & 0.3  & 0.7 & 70 & 110 & 49 & $400^{3}$ & $1.73\times 10^{9}$ &
  6.6 & 0.9 & 8.9\\
  \textbf{Baby} & 0.307  & 0.693 & 67.7 & 100 & 99 & $512^{3}$ & $6.36
  \times 10^{8}$ &
  5 & 0.829 & 4.9\\
  \textbf{Flora} & 0.307  & 0.693 & 67.7 & 2000 & 99 & $1024^{3}$ & $6.35
  \times 10^{11}$ &
  48 & 0.829 & 4.9\\
  \hline
\end{tabular}
\caption{Main features of the three simulations used in this work. \label{tab_sim}}
\end{table*}

In  this work  we  use  the results  of  three different  cosmological
simulations.  First,  the GIF2  Simulation \citep{gao04}: it  adopts a
$\Lambda$CDM     cosmological     model     with     $\Omega_{m}=0.3$,
$\Omega_{\Lambda}=0.7$,  $\sigma_{8}=0.9$  and   $h=0.7$  and  follows
$400^{3}$ particles in a periodic cube of side $110h^{-1}$ Mpc from an
initial redshift $z=49$ to the  present time.  The individual particle
mass is $1.73\times 10^{9}h^{-1}$ $M_\odot $.  Initial conditions were
produced  by  imposing perturbations  on  an  initially uniform  state
represented  by a  glass  distribution  of particles  \citep{white96}.
Based on the Zel'dovich  approximation \citep{zeldovich70}, a Gaussian
random field  is set up by  perturbing the positions of  the particles
and assigning them velocities according  to the growing model solution
of linear theory \citep{seljak96}.

Then, within a serie of new  cosmological simulations, we ran Baby and
Flora, using the publicly available code GADGET-2 \citep{springel05a}.
In particular, Baby  follows $512^{3}$ particles in a  periodic box of
$100h^{-1}$ Mpc from $z=99$ to the  present time. The particle mass is
$6.36\times  10^{8}h^{-1}$ $M_\odot  $  and adopts  $\Omega_{m}=0.307$
($\Omega_{b}  =  0.0483$  used  to  compute  the  transfer  function),
$\Omega_{\Lambda}=0.693$, $\sigma_{8}=0.829$  and $h=0.677$, following
the  recent  Planck  results  \citep{planckxvi}.   The  initial  power
spectrum was  generated with  the code  CAMB \citep{camb}  and initial
conditions were produced perturbing  a glass distribution with N-GenIC
(\url{http://www.mpa-garching.mpg.de/gadget}).       Flora     follows
$1024^{3}$ particles in a periodic box  of $2 h^{-1}$ Gpc from $z=99$,
with the  same cosmological model  of the previous one.   The particle
mass is $6.35\times  10^{11}h^{-1}$ $M_\odot$ and has  been run mainly
to have a larger statistic for massive clusters-size haloes.

\begin{figure}
\includegraphics[width=\hsize]{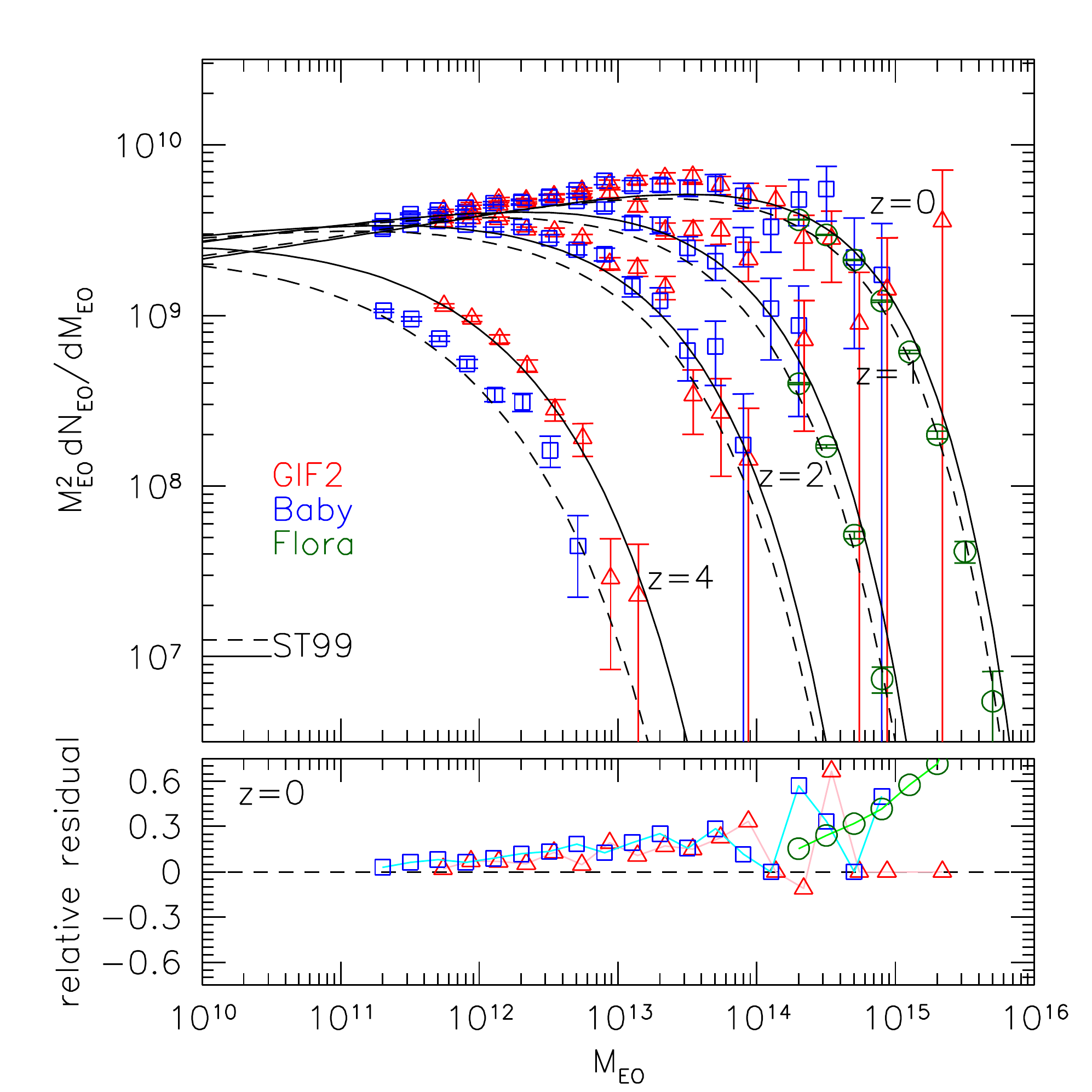}
\caption{Halo  mass  function  of  the  three  simulations  at  four
  different  $z$,  obtained  with  the  EO  halo  finder.
  Different  data points  and colors  show the  results for  the three
  different  simulations, and  the  error bars  represent the  Poisson
  uncertainty.  The  solid and dashed curves  represent, respectively,
  the theoretical prediction  for the GIF2 and  Planck cosmology given
  by \citet{sheth99b} mass  function (ST99). The lower panel show the relative residuals 
  between the halo abundance obtained with the EO and the SO finder for the three
  simulations at $z=0$.\label{massf}}
\end{figure}

The combined use of two different cosmological models will allow us to
shed more  light on the possible  dependence of the evolution  of halo
shape  and evolution  as a  function of  cosmological parameters.   In
Table~\ref{tab_sim} we  summarize the parameters with  which the three
simulations have been  run. For example, we see how  the difference in
cosmology  produces a  difference in  $M_{*}$  - the  typical mass  of
haloes  forming   at  the  present   time,  which  is   identified  by
$\nu=\nu(z)=\delta_{c}(z)/\sigma(M)=1$, where $\sigma^2(M)$ represents
the variance in the initial density fluctuation field when smoothed on
scale $R = (3M/4\pi\bar\rho)^{1/3}$  and $\delta_{c}(z)$ is the linear
overdensity threshold  for collapse extrapolated at  redshift $z$.  At
$z=0$      for      the     GIF2      simulation      $M_{*}=8.9\times
10^{12}M_{\odot}h^{-1}$,     while    for     Baby    $M_{*}=4.9\times
10^{12}M_{\odot}h^{-1}$ due to the different cosmological model.

\subsection{Post-processing pipeline}
At each simulation snapshot we  first identified dark matter haloes as
peaks in the density field adopting a Spherical Overdensity (hereafter
SO)  algorithm: we  estimated the  local  dark matter  density at  the
position of  each particle  by calculating the  distance to  the tenth
closest neighbour  and we  assigned to each  particle a  local density
$\rho \propto d_{10th}^{-3}$.  Sorting particles in density, we choose as
centre of  the first halo  the position  of the densest  particle.  We
then grow  a sphere of  matter around this  centre and stop  when the
density within the sphere falls below the virial value appropriate for
the cosmological model at that  redshift; for the definition of virial
density  we adopted  the model  of \citet{eke96}.   Properties of  all
systems with  more than $10$ particles  are stored by our  halo finder
\citep{tormen04, giocoli08b}.  In this way, for the three simulations,
we  obtain  a  catalogue  of  spherical  virialized  haloes  for  each snapshot.

Even  if  the SO  algorithm  has  been proven  to  work  very well  in
identifying haloes  and it has  also been shown that  spherical haloes
can be  used to estimate the  halo mass function  and other properties
quite precisely, it is also true that it is more realistic to describe
haloes  as   triaxial  ellipsoids,  as  within  the   context  of  the
Ellipsoidal   Collapse    (hereafter   EC)   model   (\citet{white79},
\citet{bond96}, \citet{sheth01b}).  This is motivated by the fact that
haloes are not isolated systems and that the surrounding gravitational
field  influences them  during  their collapse  and formation  phases;
moreover  during their hierarchical  growth they  experience different
merging. All these effects stretch and modify the halo shape.

Thus, an ellipsoidal halo finder is particularly useful when one wants
to study halo shapes, which are obviously more sensible than any other
properties  to the  way  in  which haloes  are  identified.  For  this
reason,  after the  SO algorithm  we ran  the  Ellipsoidal Overdensity
(hereafter EO)  code described in  \citet{despali13}: it re-identifies
all haloes  previously found  by the SO  algorithm using  an iterative
method  to   obtain  the  best  fitting  ellipsoid   to  the  particle
distribution;  still,  halo boundaries  are  chosen  using the  virial
overdensity.   Note that, in principle, the EO algorithm could
run on any pre-existing halo catalogue: we chose to use a combination
of SO+EO algorithms to be more consistent, since both method define
haloes using a particular shape and since we
need to grow a sphere as a first step of the EO (for the starting
guess on the shape). We then  calculated the  mass  tensor $M_{\alpha\beta}$
defined by the $N$ particles found inside the ellipsoid as:
\begin{equation}
 M_{\alpha\beta} = \frac{1}{N}\sum_{i=1}^{N} r_{i,\alpha}r_{i,\beta}
\end{equation}
where $\textbf{r}_{i}$  is the position  vector of the  $i$th particle
and  $\alpha$ and $\beta$  are the  tensor indices.   By diagonalizing
$M_{\alpha\beta}$  we   then  obtained  the   eigenvalues  ($l_{1}\geq
l_{2}\geq  l_{3}$)  and  eigenvectors   which  define  the  shape  and
orientation     of    the     virialized    structure:     the    axes
($\lambda_{1}\geq\lambda_{2}\geq\lambda_{3}$)  of   the  best  fitting
ellipsoid  are  defined  as  the  square  roots  of  the  mass  tensor
eigenvalues:  $\lambda_{i}=\sqrt{l_{i}}$.   In  this way  we  obtained
another set  of EO catalogues, equivalent  to the SO one:  we chose to
keep only haloes formed by more  than $200$ particles to ensure a good
resolution  in   the  determination   of  the  axes.    The  resulting
ellipsoidal  masses  are  slightly  higher  than  the  spherical  ones
(consistently  to the  fact that  ellipsoids are  expected to  fit the
actual  density distribution  better than  spheres) and  the resulting
shapes  are obviously  more  elongated \citep{despali13}.  In
  Figure  \ref{massf}  we show  the  resulting  ellipsoidal halo  mass
  function  at  four  different  $z$  for the  three  simulations;  in
  particular for Flora we show  only two redshifts since there are not
  many  haloes resolved  with more  than $200$  particles  at redshift
  larger than  one; the error bars represent  the Poisson uncertainty.
  Solid and dashed line represent the halo mass function prediction by
  \citet{sheth99b} for the two cosmologies of the simulations.  In the
  lower  panel,  we show  the  relative  residual  between the  mass
  functions  obtained  with  the  EO  and SO  finders  at  $z=0$  (the
  behaviour at the other redshift  is consistent with this one). Since
  the ellipsoidal and the  spherical masses are positively biased with
  respect to  each other of some  percents, this results in  a bias in
  the the halo  abundance that can reach $10-30\%$  and that increases
  with halo mass due to the fact that massive haloes are intrinsically
  more elongated and so their  mass and shape are captured much better
  by an ellipsoidal shape.

\subsection{Halo shape parameters}
As explained  in the previous  section, the  axes of the  best fitting
ellipsoid for  each halo are defined  by the square roots  of the mass
tensor eigenvalues. The other relevant  quantities that can be derived
from the  eigenvalues are the  ellipticity $e$ and prolateness  $p$ of
each halo, that can be written as:
\begin{equation}
e= \frac{\lambda_{1}-\lambda_{3}}{2\tau} \qquad \mathrm{,} \qquad p =
\frac{\lambda_{1}-2\lambda_{2}+\lambda_{3}}{2\tau} \label{e-p}
\end{equation}
where  $\lambda_{1,2,3}$ are  the axes  derived from  the mass  tensor
eigenvalues  and  $\tau=\lambda_{1}+\lambda_{2}+\lambda_{3}$ .   Using
this definitions,  $e$ quantifies  the deviations from  sphericity and
$p$ measures  prolateness versus oblateness: by  construction a sphere
has            $e=p=0$.            The            fact            that
$\lambda_{1}\geq\lambda_{2}\geq\lambda_{3}\geq   0$   introduces   the
boundaries $-e\leq p  \leq e$, $p\geq 3e-1$. This makes  the points of
the $e-p$ distribution  populate only a triangle, as  already shown in
\citet{bardeen86},\citet{porciani02b}  and  \citet{desjacques08},  and
here in  Figure \ref{ellpro}  which we  will discuss  later on  in the
paper.

\section{Results: halo populations at different $z$}\label{results1}
In  this  section  we  will  describe how  the  halo  ellipticity  and
prolateness  change as  a  function of  redshifts  for different  halo
masses and how it is possible to obtain universal relations for the
shape parameters.

\subsection{Distributions of $e$ and $p$ at different times}

\begin{figure*}
\includegraphics[width=\hsize]{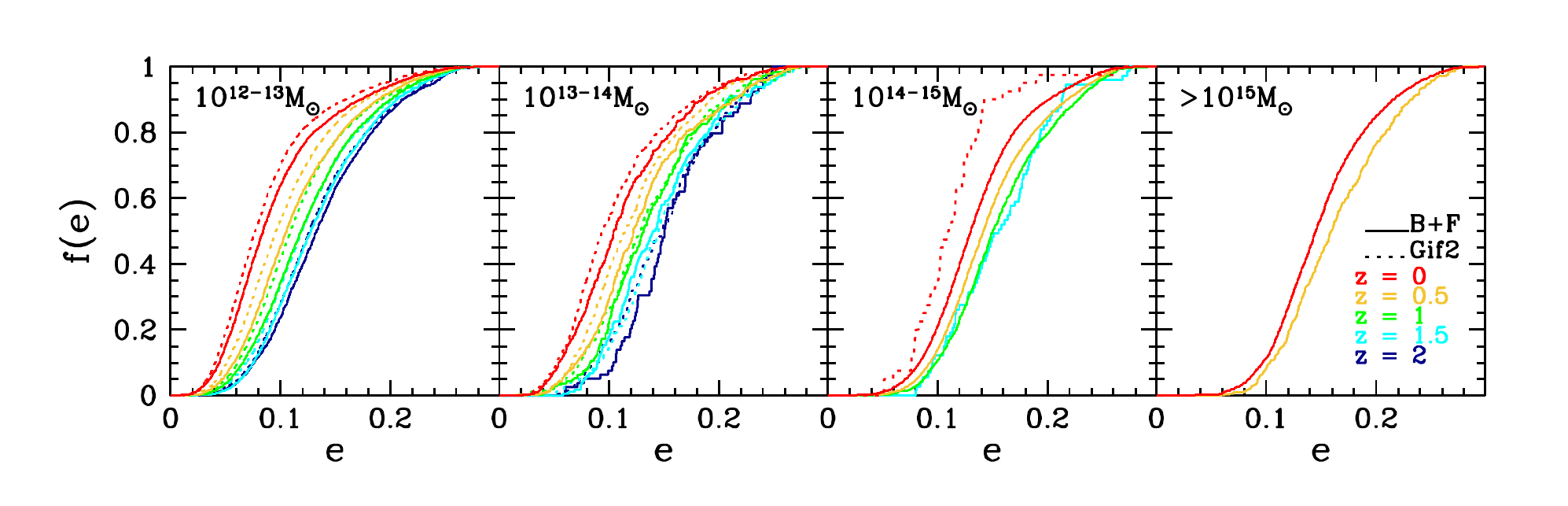}
\includegraphics[width=\hsize]{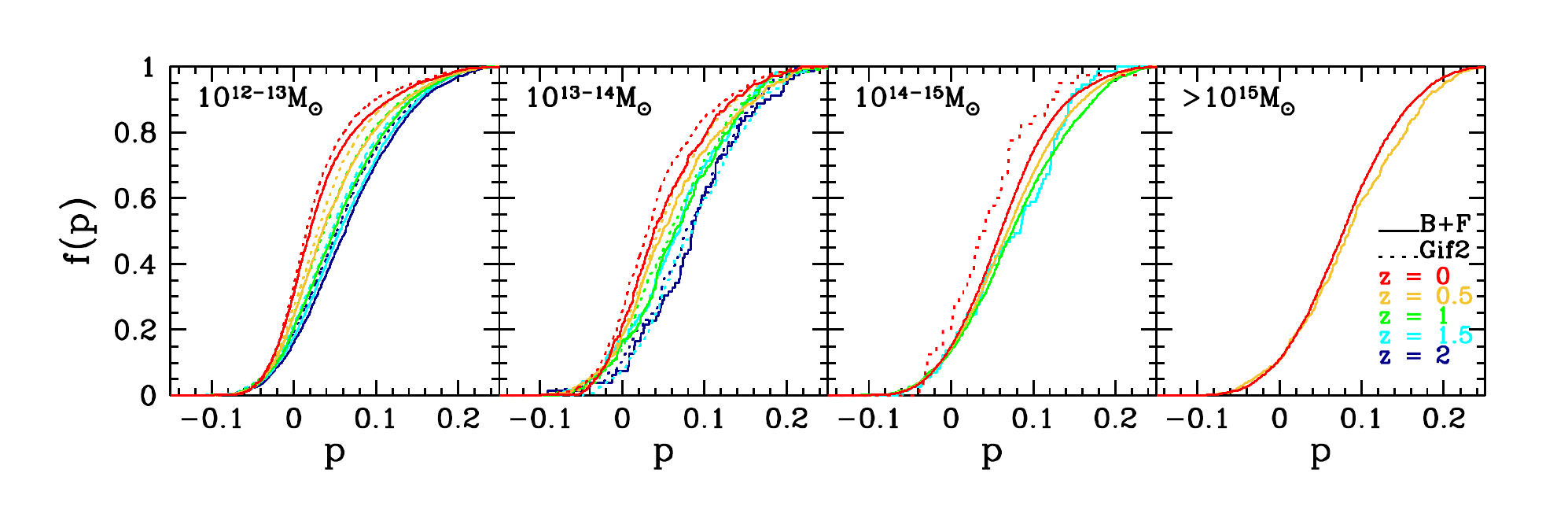}
\caption{Ellipticity  and prolateness  cumulative distributions.  Each
  panel shows the distribution at  five redshifts for a different mass
  bin, with increasing mass from left to right. The haloes of Baby and
  Flora are represented by the solid lines, while those of the GIF2 by
  the dotted  lines. We notice  that both ellipticity  and prolateness
  decrease     to    low     redshift     and     also    to     lower
  masses.\label{cumulatives}}
\end{figure*}
\begin{figure*}
\includegraphics[width=7cm]{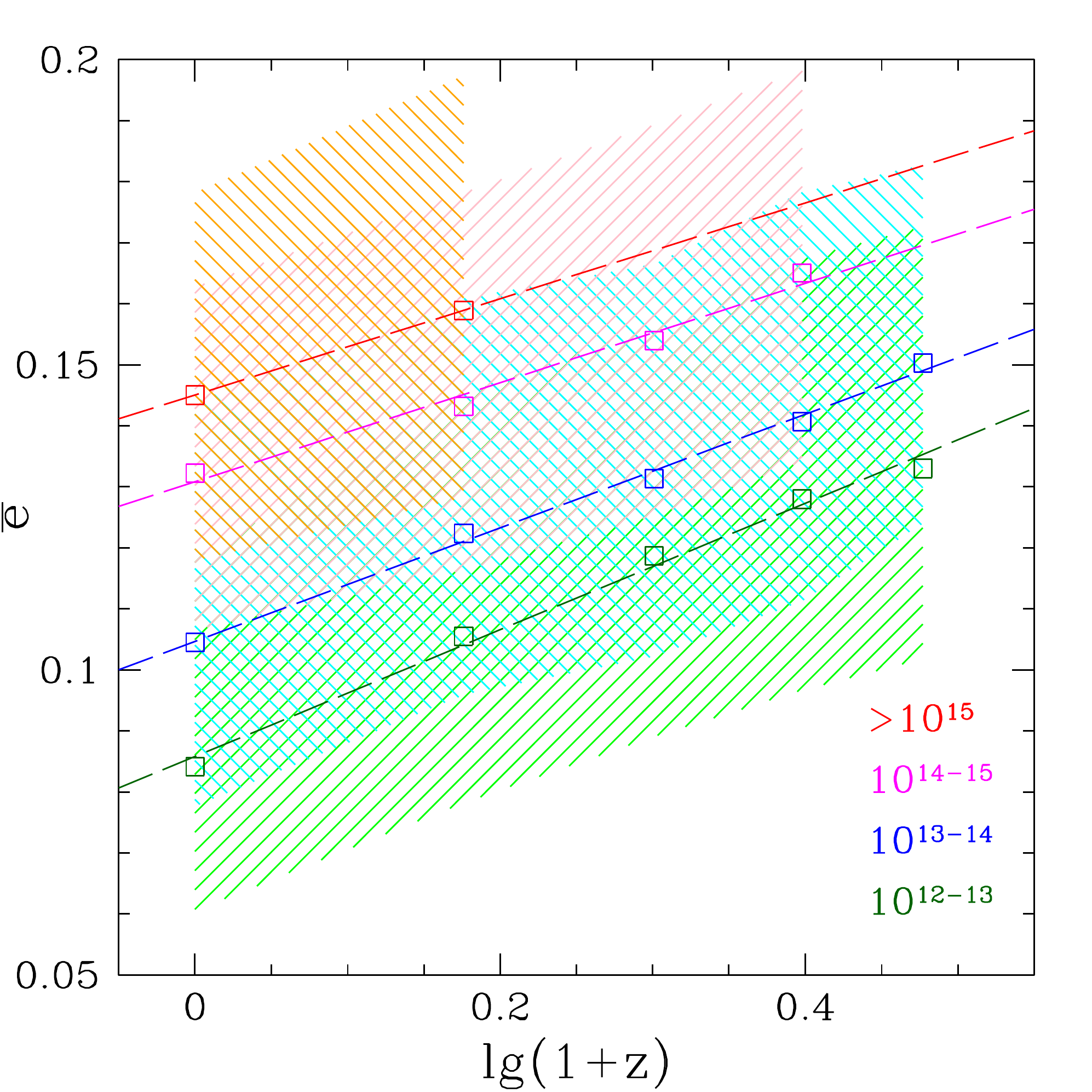}
\includegraphics[width=7cm]{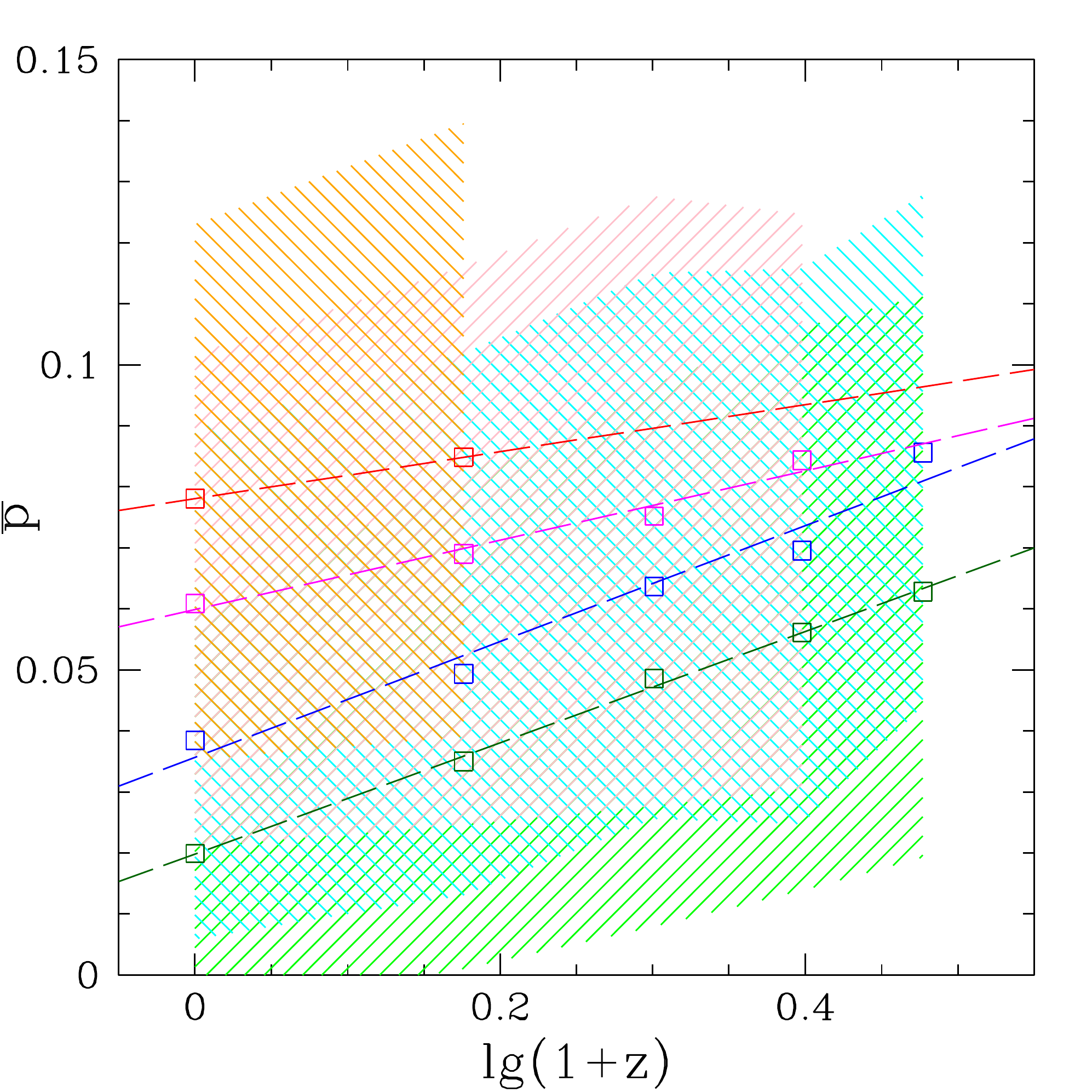}
\caption{Medians  and quartiles  of  the  cumulative distributions  of
  Figure \ref{cumulatives} both for  the ellipticity (left panel) and
  the prolateness  (right panel), for Baby  and Flora as a  function of
  redshift.  Each  set  of  point  shows the  median  relation  for  a
  different mass bin and the  shaded regions the quartiles. The dashed
  lines    represent    the    linear    fit   to    each    set    of
  points.\label{cum_med}}
\end{figure*}

In  Figure   \ref{cumulatives}  we   show  the  ellipticity   and  the
prolateness  cumulative distributions:  in each  panel we  present the
results for five different redshifts  - $z=2,1.5,1,.5,0$ - (or less at
high masses) for  a given mass bin.  The haloes of  Baby and Flora are
represented together  - since  they have the  same cosmology -  by the
solid lines,  while those of the  GIF2 by the dotted  lines.  Flora is
determinant  to have  enough data  at $M>10^{14}  M_{\odot}h^{-1}$: it
contains around  150000 systems more  massive than $10^{14}$  and 1390
still more  massive than  $10^{15} M_{\odot}h^{-1}$.  Looking  at each
panel it is clear how  both ellipticity and prolateness peak at higher
values at  high redshift; on the  other hand, comparing  the curves of
the  same  colour in  the  four  panels, we  see  that,  at any  given
redshift, the  median of $e$ and  $p$ increases with  mass: as already
shown     in     other     works     using    the     axial     ratios
\citep{allgood06,munozcuartas11}, at  the present time --  and also at
each previous  epoch --  the most massive  systems are also  the least
spherical.  They are  still in the formation phase  and so their shape
is still be influenced by the direction of the last major merger or of
the  material  falling in  through  the  filaments,  making them  more
elongated. Smaller  haloes, formed  at higher redshifts  and typically
more concentrated, lived  for enough time to relax  and lose memory of
the  directions of  the  different merging  events experienced  during
their  history.    To  stress   these  two  dependencies,   in  Figure
\ref{cum_med}, we show, as a function of $z$ -- only for Baby+Flora --
the medians of the ellipticity and of the prolateness distributions of
Figure \ref{cumulatives}: each set  of point shows the median relation
for a  different mass  bin and the  shaded regions the  quartiles; the
dashed lines represent the linear fit to each set of points.

 In Figure \ref{cumulatives},  we observe also a  slight dependence on
 cosmology, with  Baby and  Flora having an  higher average  value for
 both $e$ and $p$; this was expected since, in a universe with a lower
 value of $\sigma_{8}$ (Baby+Flora), haloes tend to form later and so,
 when we look at them at a given time, they are still more ellipsoidal
 than those which form with in an higher-$\sigma_{8}$-universe (GIF2).

\subsection{A universal rescaling for halo shape evolution}
\begin{figure*}
\includegraphics[width=0.85\hsize]{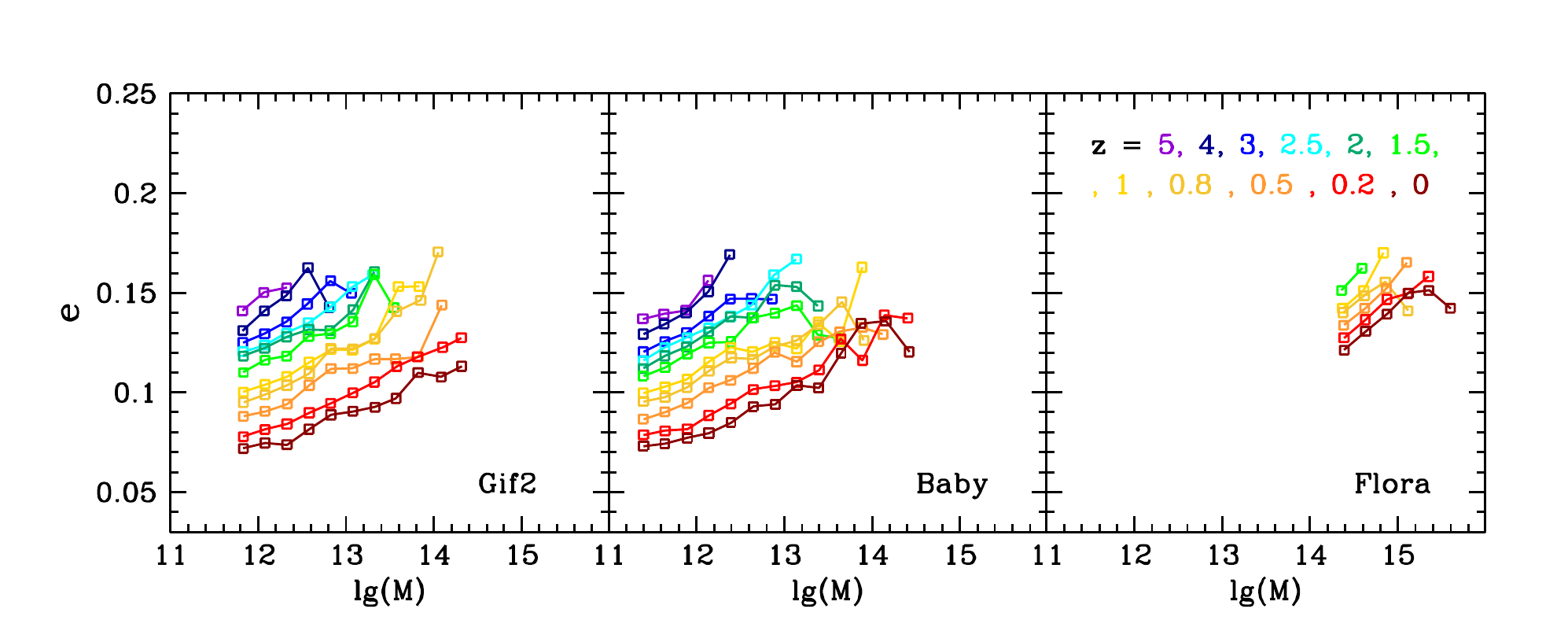}
\includegraphics[width=0.85\hsize]{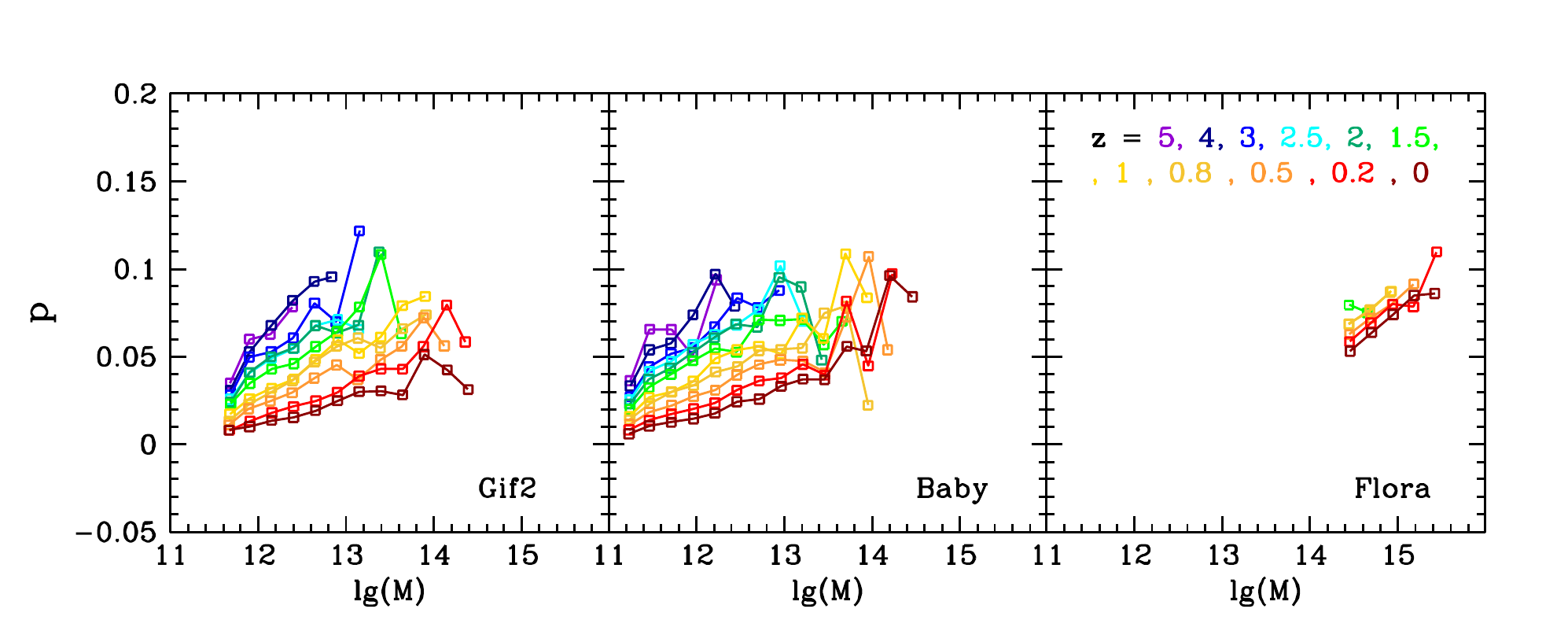}
\caption{Ellipticity   ($top$)  and   prolateness   ($bottom$)  median
  distributions  as  a  function  of  mass  for  13  outputs  of  both
  simulations: the  three panels refer respectively to  the GIF2, Baby
  and Flora simulations.\label{ellpro_mass_3sim} }
\end{figure*}

In  Figure  \ref{ellpro_mass_3sim} we  show  how  the ellipticity  and
prolateness  distributions  evolve in  time  for  haloes of  different
masses: in the top panel the median ellipticity is plotted against the
halo mass for eleven snapshots  of each simulation (a part from Flora,
we have a significant number of haloes only at six snapshots).  We can
notice  both  a   dependence  on  mass  and  on   time  as  in  Figure
\ref{cumulatives}: first, looking at each set of points independently,
it is  clear that  more massive haloes  are on average  less spherical
than the  smaller ones.  Then, looking  at the whole  plot, the median
ellipticity   decreases  in   time,  leading   to  a   more  spherical
distribution at  the present  time. In the  bottom panel, we  show the
same results for prolateness.  At  all times haloes tend to be prolate
($p>0$), even if  this trend weakens at low  redshifts.

\begin{figure*}
\includegraphics[width=0.85\hsize]{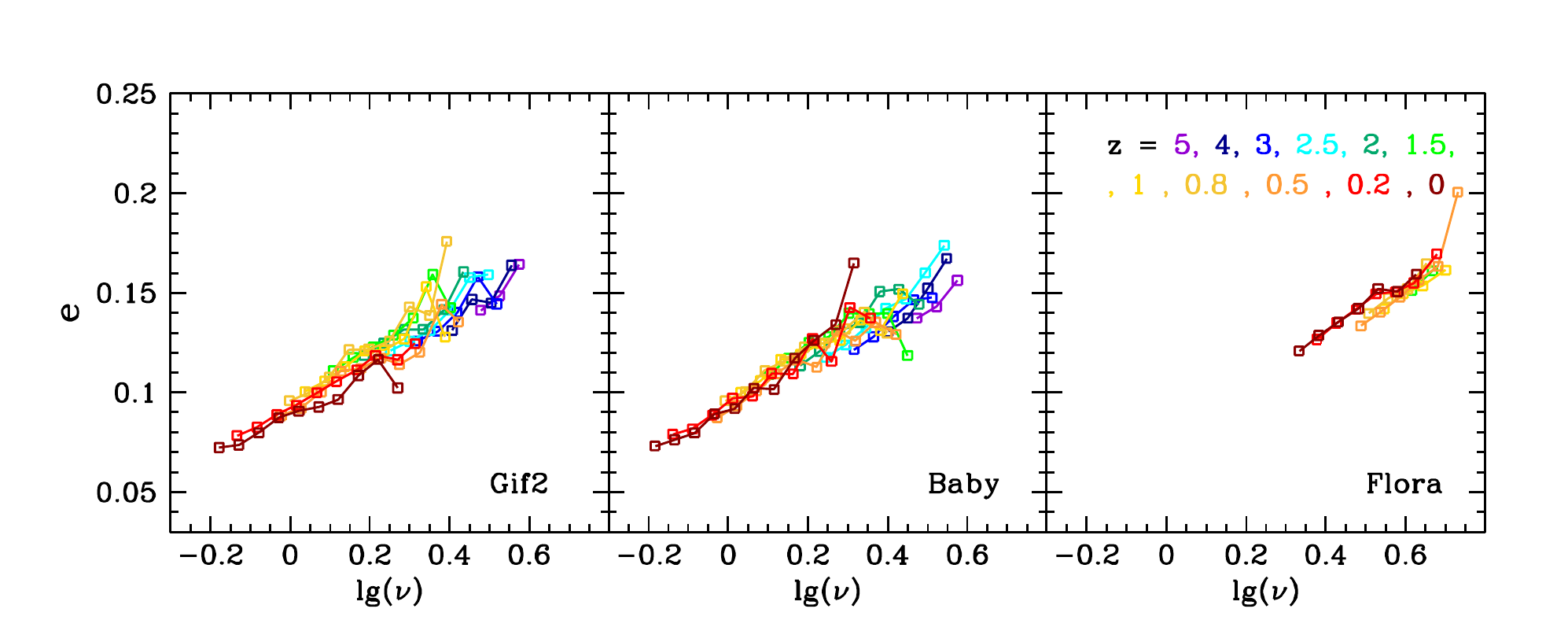}
\caption{Median Ellipticity as a function of $\nu$: we show the effect
  of rescaling the  mass to the variable $\nu$: since  it contains the
  dependence on epoch and cosmology, all the ellipticity distributions
  of  the previous  Figure now  lie on  the same  relation. The  color
  scheme is  the same of Figure  \ref{ellpro_mass_3sim}. The rescaling
  has  also the  same  effect  on the  prolateness  and  on the  axial
  ratios.\label{ell_rescaling}}
\end{figure*}

\begin{figure*}
\includegraphics[width=0.4\hsize]{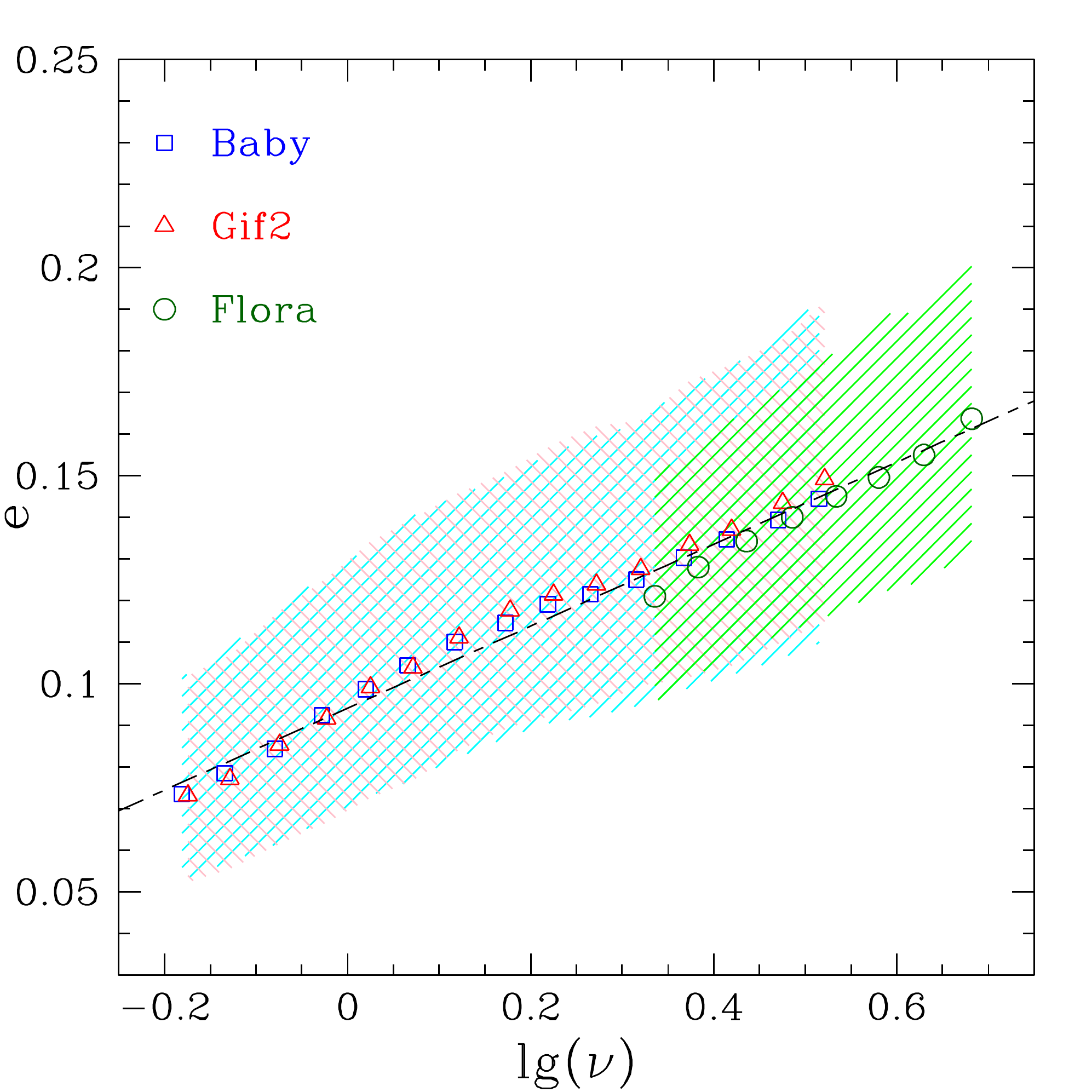}
\includegraphics[width=0.4\hsize]{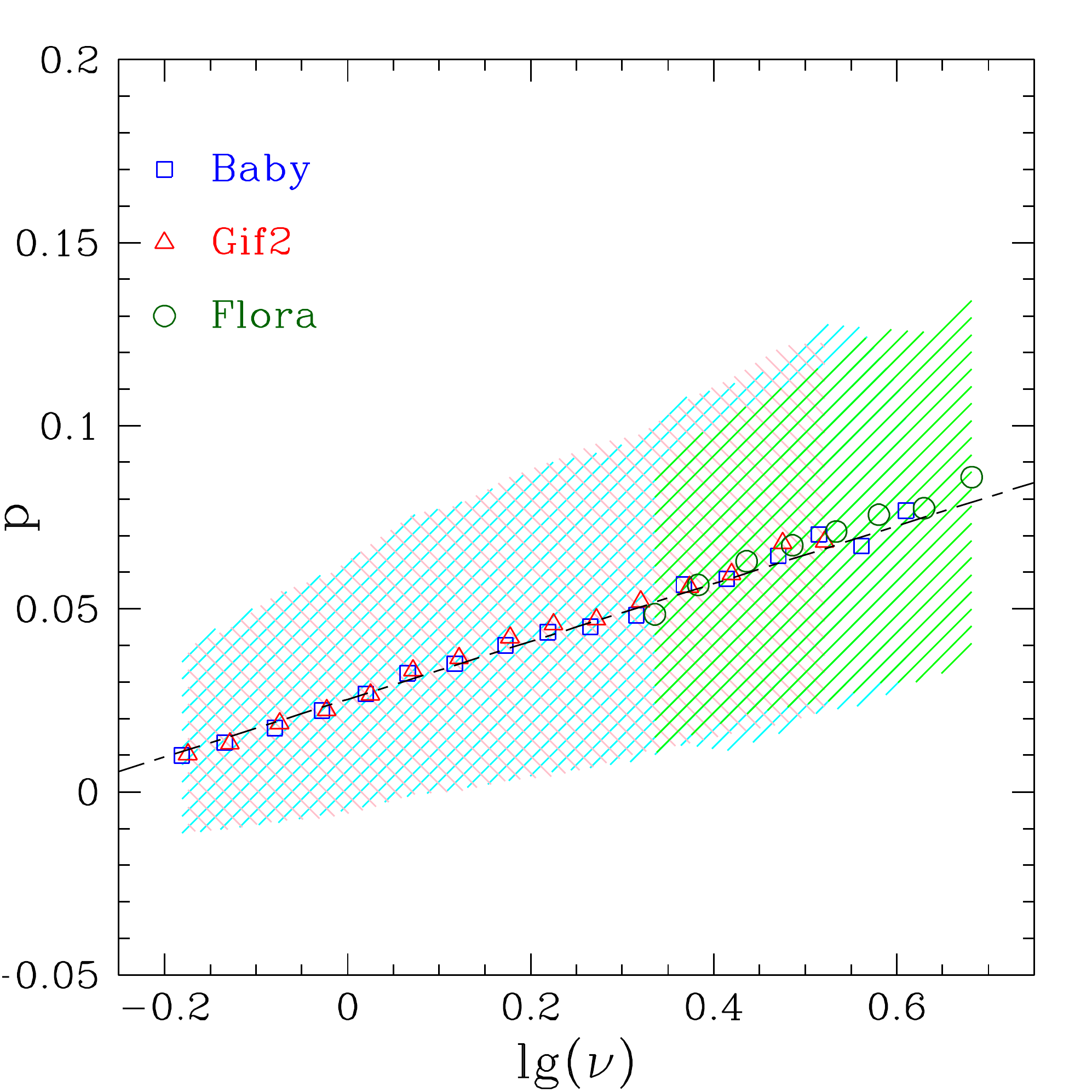}
\caption{$Universal$ ellipticity and prolateness distribution. $e$ and
  $p$ are shown as a function of the variable $\nu=\delta_{c}/\sigma$:
  this choice eliminates  the dependence on epoch and, as  we see, the
  distributions  at all  times  lies on  same  relation. The  coloured
  points show  the medians for  each simulation and  the corresponding
  shaded coloured regions enclose the first and third quartile of each
  distribution; the black  dashed line represents the best  fit to all
  the points.\label{ellpro_nu_3sim}}
\end{figure*}

Since  the   virial  mass   is  a  cosmology   and  redshift-dependent
definition,  Press  and  Schechter  (PS)  and  extended-PS  approaches
\citep{press74,bond91,lacey93}   have    shown   that   an   universal
generalisation  of the  mass function  can  be obtained  by using  the
variable  $\nu(z)$.   An analogous  approach  has  also  been used  by
\citet{prada11} to rescale the concentration for different cosmologies
and various  redshifts. Also in our  case, using $\nu$  instead of the
virial  mass,  allows  to  obtain  an universal  relation:  in  Figure
\ref{ell_rescaling}    we   show   the    same   points    of   Figure
\ref{ellpro_mass_3sim}, but  as a function  of $\nu$ instead  of mass.
It is easy to see that the  use on $\nu$ remove the dependence on both
cosmology and time:  this is due to the  fact that $\sigma(M)$ retains
the  information of  the  mass and  is  higher for  low masses,  while
$\delta_{c}(z)$  changes  in  time,   increasing  at  high  $z$.   The
combination of  the three  simulations allows us  to span an  order of
magnitude in $\nu$  and we verify that, in these  unit, all the points
move on the  same median relation. The effect is the  same also on the
prolateness and  the axial  ratios and  so we believe  that it  is not
worth showing  all of them.   Thus, in Figure  \ref{ellpro_nu_3sim} we
decided to  average over all the  points at all  the eleven redshifts,
for each one of the  simulations: the coloured points show the medians
for each simulation and  the corresponding shaded regions enclose the
first and third quartile of  each distribution.  The black dashed line
represents the  best fit to all  the points, which can be written as:
\begin{eqnarray}
e = 0.098^{+0.001}_{-0.001}\lg(\nu) +0.0940^{+0.0002}_{-0.0001}
\nonumber \\
p = 0.079^{+0.003}_{-0.003}\lg(\nu)+ 0.025^{+0.001}_{-0.001} .
\end{eqnarray}
The  parameters   and  the  errors  were  obtaining   by  fitting  the
distributions in both directions and then taking the mean values.  For
$e$,  the interquartile  difference goes  from 0.05  al low  values of
$\nu$ to  0.08 at  high $\nu$; for  $p$ it  changes from 0.05  to 0.1.
This  reflects the fact  that haloes  at different  redshifts populate
different regions due  to their relative mass limits,  but they remain
around it.

\begin{figure*}
\includegraphics[width=0.4\hsize]{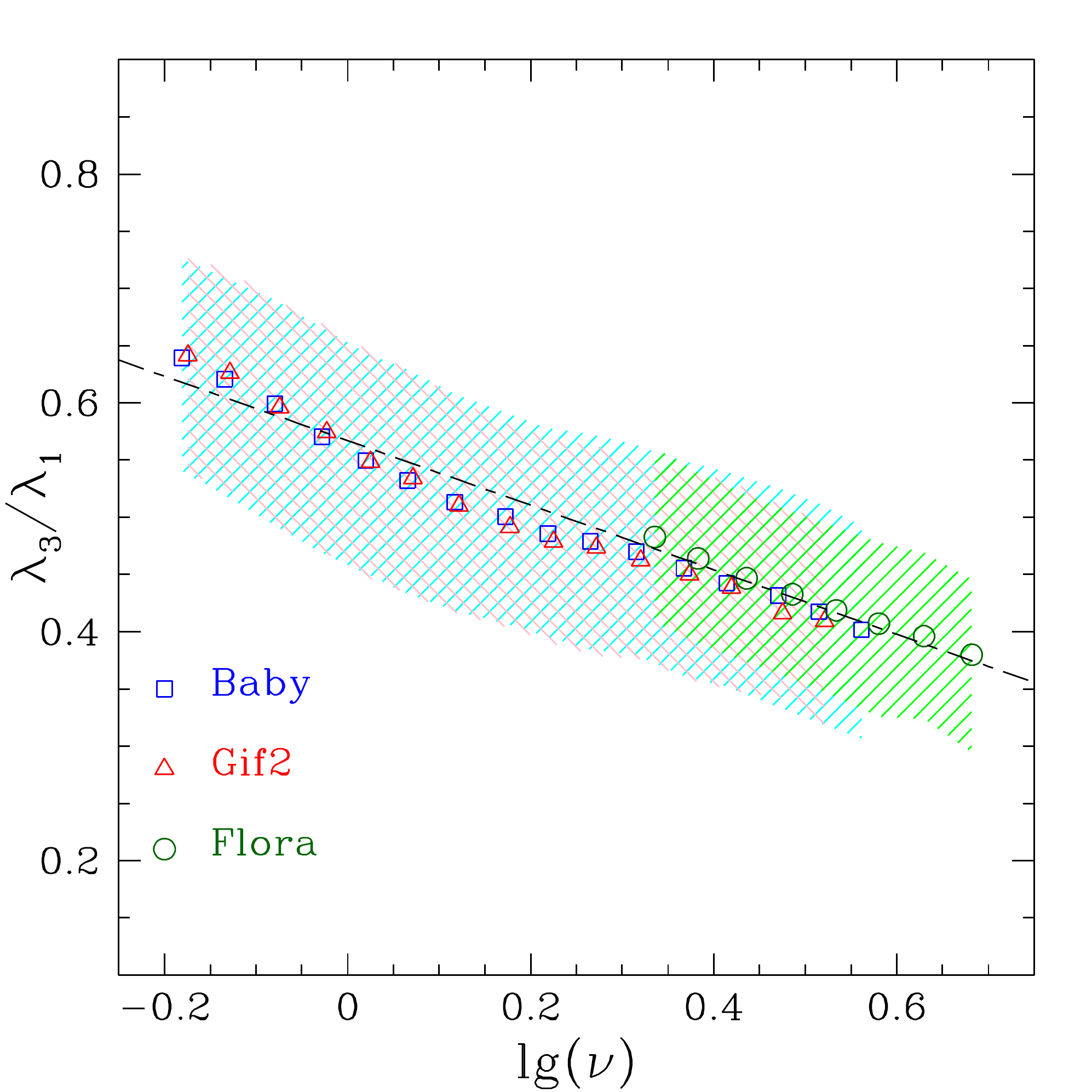}
\includegraphics[width=0.4\hsize]{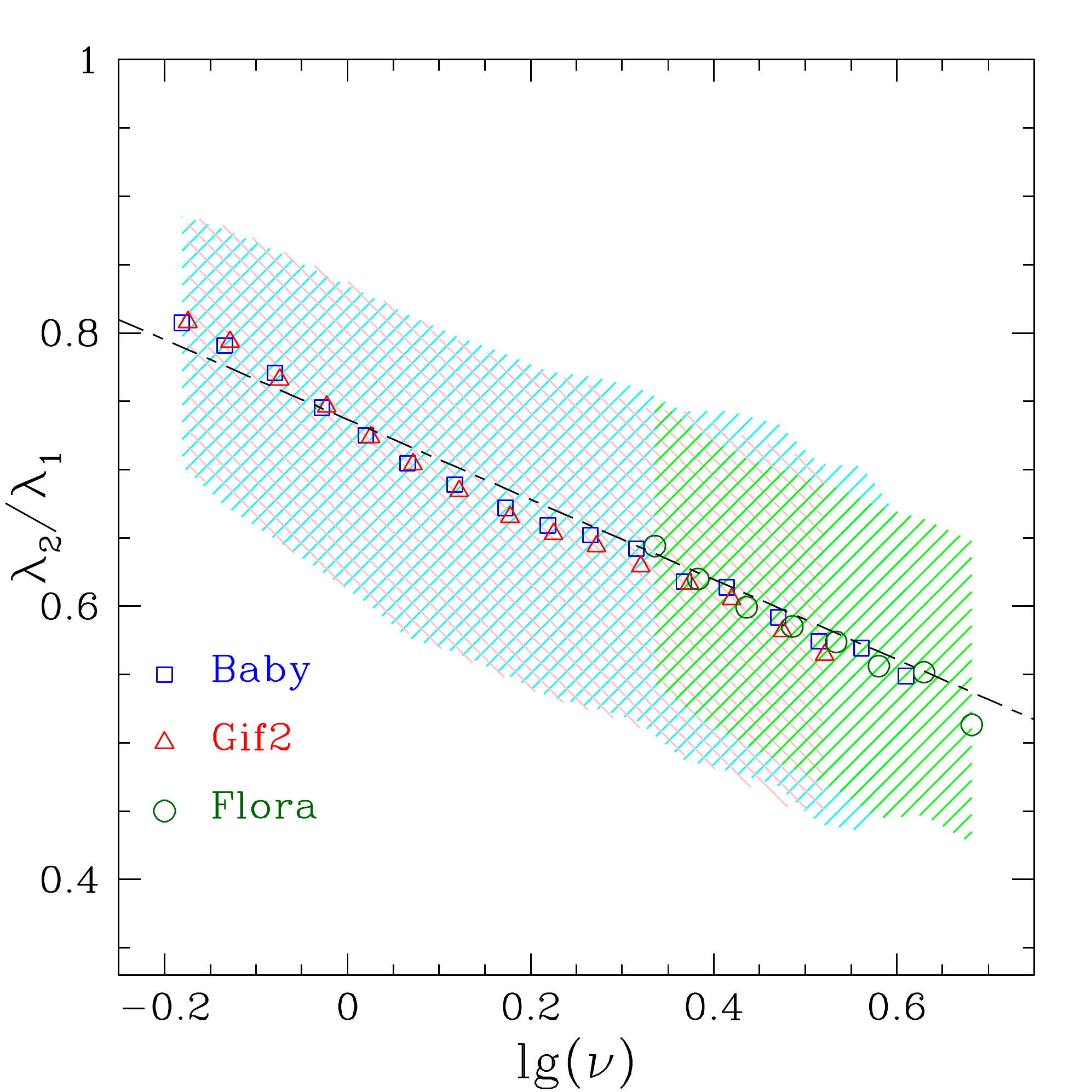}
\caption{ $Universal$  axial ratio distributions. The  color scheme is
  the same as in Figure \ref{ellpro_nu_3sim} \label{axial_ratios}}
\end{figure*}

In Figure  \ref{axial_ratios} we show  the same results for  the axial
ratios   ($\lambda_{3}/\lambda_{1}$   and  $\lambda_{2}/\lambda_{1}$),
which  can be  more useful for the  comparison with
observations. Looking at the axial ratios we recognise the same
trends in the evolution of shapes that we noticed studying $e$ and
$p$. The best fit relations to all the data points are:
\begin{eqnarray}
\frac{\lambda_{3}}{\lambda_{1}} = -0.282^{+0.003}_{-0.004}\lg(\nu) +0.567^{+0.001}_{-0.001} \nonumber \\
 \frac{\lambda_{2}}{\lambda_{1}}  = -0.293^{+0.007}_{-0.005}\lg(\nu)+ 0.736^{+0.002}_{-0.002}.
\end{eqnarray}
In this case the interquartile difference for
$\lambda_{3}/\lambda_{1}$ goes from 0.17 at low $\nu$ to 0.19 at high
ones; for $\lambda_{2}/\lambda_{1}$ it changes more, from 0.17 to 0.25. 

\subsection{$e$-$p$ relation}
\begin{figure}
\includegraphics[width=\hsize]{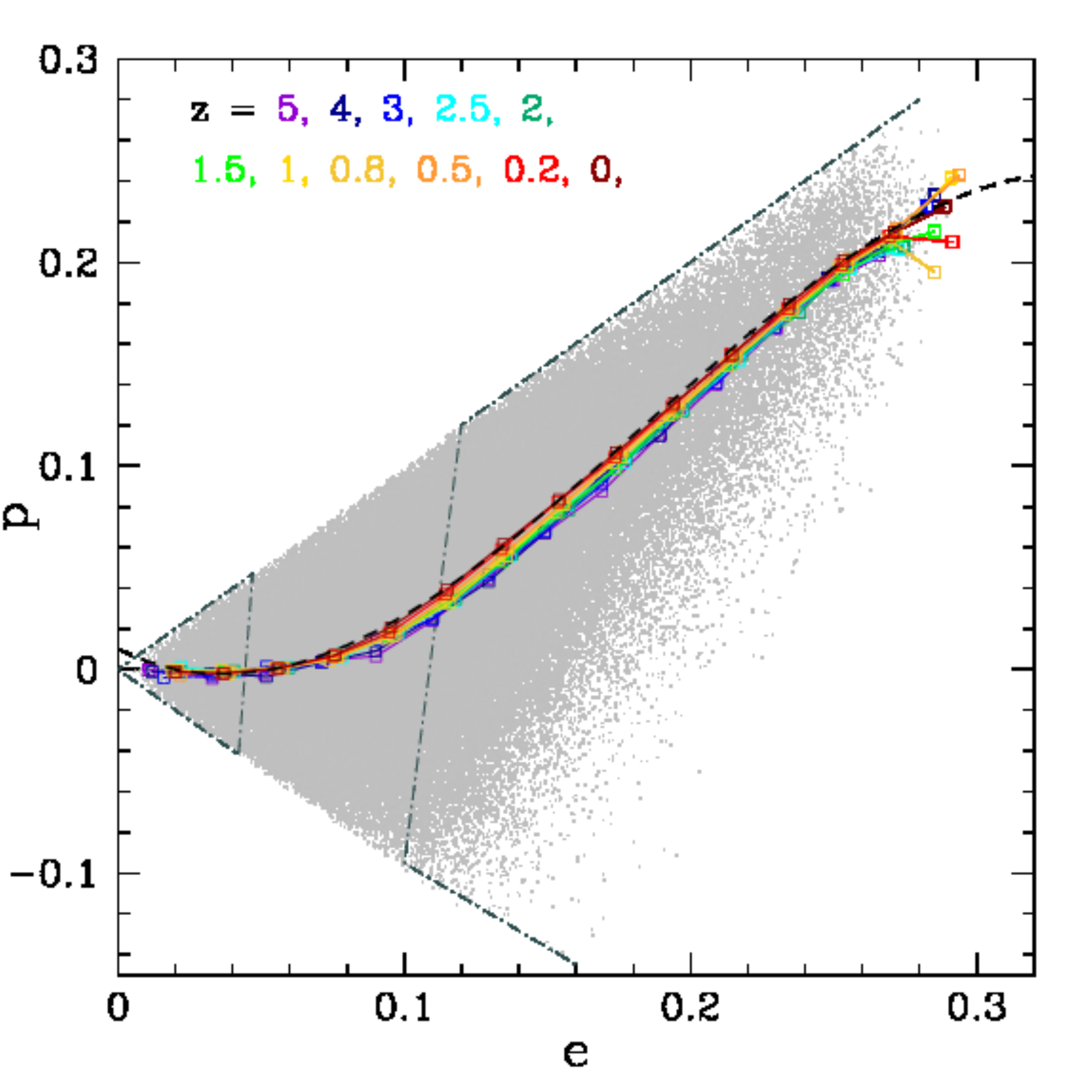}
\caption{$e$-$p$ distribution at eleven  different redshifts, the same
  of  Figure  \ref{ellpro_mass_3sim}:  the coloured  points  show  the
  medians at  a certain redshift  taken on all  the data of  the three
  simulations  together.  The  black  dashed curve  shows the  fitting
  function to the  points at $z=0$. The gray dots  represent the $e-p$
  distribution  at $z=0$  for all  the simulations;  the dotted  lines
  delimit  the  regions  for  which the  relative  difference  between
  $\lambda_{1}$ and $\lambda_{3}$ is less than $25\%$ (in the triangle
  on  the   left)  or  more  than   $50\%$  (in  the  region   on  the
  right).\label{ellpro}}
\end{figure}

As    discussed   in   \citet{bardeen86},    \citet{porciani02b}   and
\citet{desjacques08}, the definition of $e$ and $p$, together with the
range  of the eigenvalues,  introduces a  correlation between  them at
high ellipticities  (which in  particular it has  been studied  at the
initial   conditions).   In   what  follows,   we  present   the  same
distribution  but using the  catalogues of  virialized haloes  at each
$z$. The ellipticity and prolateness of haloes still form a triangular
region  in the  $e-p$  plane and  it  is interesting  that the  median
distribution almost  does not change  with time (Figure~\ref{ellpro}):
at  high  redshift,  due  to   the  limited  mass  resolution  of  the
simulation, we  have few  haloes in our  mass range,  but nevertheless
they already  form a triangle; moving  to lower redshift  we have more
and more  haloes, which  keep populating the  triangle, but  leave the
median  relation   almost  unchanged.   The  data  points   in  Figure
\ref{ellpro}  show  the   median  distributions  at  eleven  different
redshifts, represented by the  different color points.  The small gray
dots  show  the  whole  distribution  at  $z=0$,  for  all  the  three
simulations together. The points at $z=0$ are fitted by the relation:
\begin{equation}
 p = 0.01-0.7e+10.57e^{2}-19.1e^{3}.
\end{equation}
which is represented by the black  dashed line; to fit the relation at
the  other snapshots it  is enough  to introduce  a the  dependence on
redshift of the  order of $(1+z)^{-0.05}$.  The flat  initial part, up
to  the fourth  median point,  corresponds  to haloes  for which  both
relative         differences          between         the         axes
($(\lambda_{1}-\lambda_{3}/\lambda_{1})$                            and
$(\lambda_{1}-\lambda_{2}/\lambda_{1})$)  are less than  $25\%$, while
the  linear  growth at  $e\geq  0.1$  is  represented by  haloes  with
$(\lambda_{1}-\lambda_{3}/\lambda_{1})$  greater  than $50\%$.   These
two regions  are marked in the  figure by the dotted  lines. This also
confirms the tendency to prolateness  already shown in other works: on
average, the  second axis  $\lambda_{2}$ is not  large enough  to have
$p<0$.   The slope of  the linearly  growing part  is close  to unity:
assuming to  neglect $\lambda_{3}$ as a first  approximation, both $e$
and $p$  depend primarily on  $\lambda_{1}$, which is larger  for high
values of  $e$, but $p$  is lowered a  bit by the contribution  of the
second axis, as can be seen in Equation \ref{e-p}.

This quasi-universality  of the $e-p$  relation is useful  to estimate
the shape properties of the  halo population at a certain redshift and
to create mock catalogues performing Monte Carlo realisations

\section{Merger Tree and Formation Redshift}\label{mergerzf}
\begin{figure*}
\includegraphics[width=0.8\hsize]{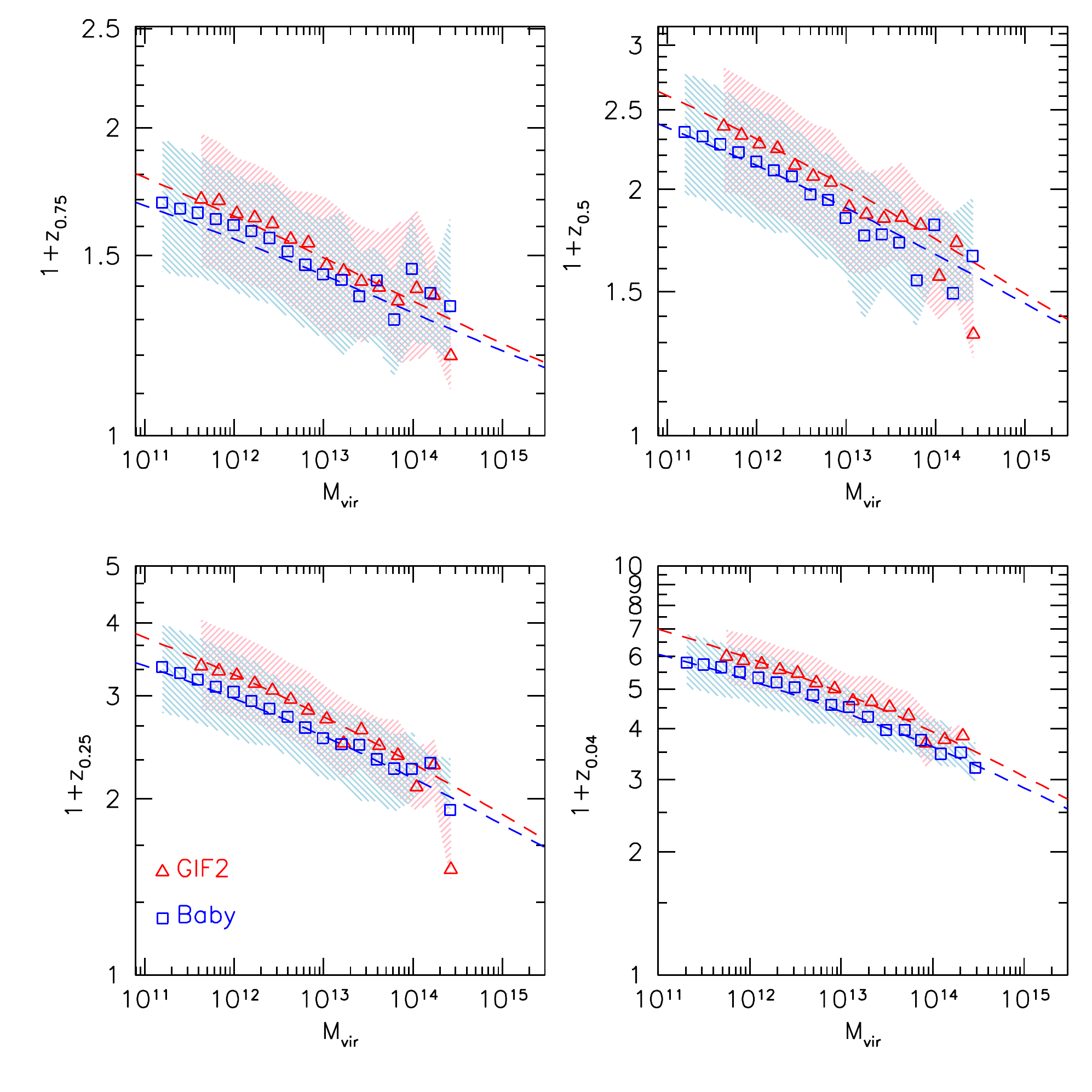}
\caption{Formation redshift  as a function  of the halo mass.   In the
  different panels we show the results derived from four definition of
  formation redshift $z_{f}$, defined as  the moment at which the main
  halo progenitor assembles a fraction $f=0.75,\,0.5,\,0.25,\,0.04$ of
  its  mass.  The  data points  show  the median  of the  measurements
  performed on  the two simulations  while the shaded  regions enclose
  the  first and  third  quartiles. The  dashed  curves represent  the
  predictions for the  formation redshift mass relations,  for the two
  cosmologies,          using           the          model          by
  \citet{giocoli12b}. \label{fmergertree}}
\end{figure*}

From the  halo catalogues, we built  the merging history tree  for all
haloes in  the simulations  with more  than $200$  particles: starting
from each  halo at redshift  $z=0$, we  define its progenitors  at the
previous output  $z=z_{i}$ as all  the haloes containing at  least one
particle of the $z=0$ halo; we term as "main progenitor" the halo that
provided  the largest  mass contribution  to  the final  one. Then  we
repeat the same procedure, now  starting from the main halo progenitor
at the snapshot $z=z_{i}$ and going backwards in time in this way from
snapshot to  snapshot, until all the  particles are lost in  the field
(i.e.   the  main halo  progenitor  possesses  fewer than  $10$
  particles). We  stress that  our approach to  follow the
  main  halo progenitor  back  in time  until it  has  fewer than  $10$
  particles is in agreement with previous works and theoretical models
  developed   to    interpret   the    halo   mass    growh   history
\citep{vandenbosch02,wechsler02,giocoli13}.          In         Figure
\ref{fmergertree} we show  the fit to the  formation redshift proposed
by \citet{giocoli12b} for the Baby  and the GIF2 cosmologies, given by
the equation:
\begin{equation}
\delta_c(z_f) = \delta_c(z_0) + \bar{w}_f \sqrt{S(f\,M) - S(M)}\,,
\end{equation}
where $z_f$ is obtained by inverting the relation between $\delta_c$
and $z_f$.
The parameters are
\begin{equation}
\bar{w}_f  = \sqrt{2\ln(\alpha_f + 1)}\,,
\end{equation}
and 
\begin{equation}
\alpha_f  = \alpha_0\,\exp(-2f^3)/f^{0.707}\,
\end{equation}
where  $\alpha_0=0.937$ --  corresponding  dashed curves  for the  two
cosmolgies.  The  value of $\alpha_0$  in this case is  different from
the one computed by \citet{giocoli12b}  by circa $15\%$ because of the
different halo sample considered.  While \citet{giocoli12b} considered
all bounds haloes that never exceed more than $10\%$ their present-day
mass along their mass accretion history,  in this work we consider all
identified systems at the present  time.  A higher value of $\alpha_0$
modifies  the normalization  of the  formation redshift-mass  relation
mainly  for large  value of  $f$  in order  to take  into account  the
accretion histories  of haloes  characterized by major  merging events
exluded by \citet{giocoli12b}. In each panel the data points represent
the median  formation redshift $z_{f}$ at  the time at which  the main
halo progenitor assembles a fraction  $f$ of its present-day mass; the
shaded regions of  the corresponding colour enclose the  first and the
third quartiles.   From the figure  we note that, since  the cosmology
adopted for  Baby (and Flora) has  a lower value of  $\sigma_{8}$, its
haloes have typically a lower  formation redshift.  Going from the top
left panel to  the bottom left one, thus decreasing  the value of $f$,
the  difference between  the two  simulations increases  up to  almost
$25\%$ for the  redshift at which the main halo  progenitor assemble a
fraction $f=0.04$ of its present-day mass.

\begin{figure}
\includegraphics[width=0.95\hsize]{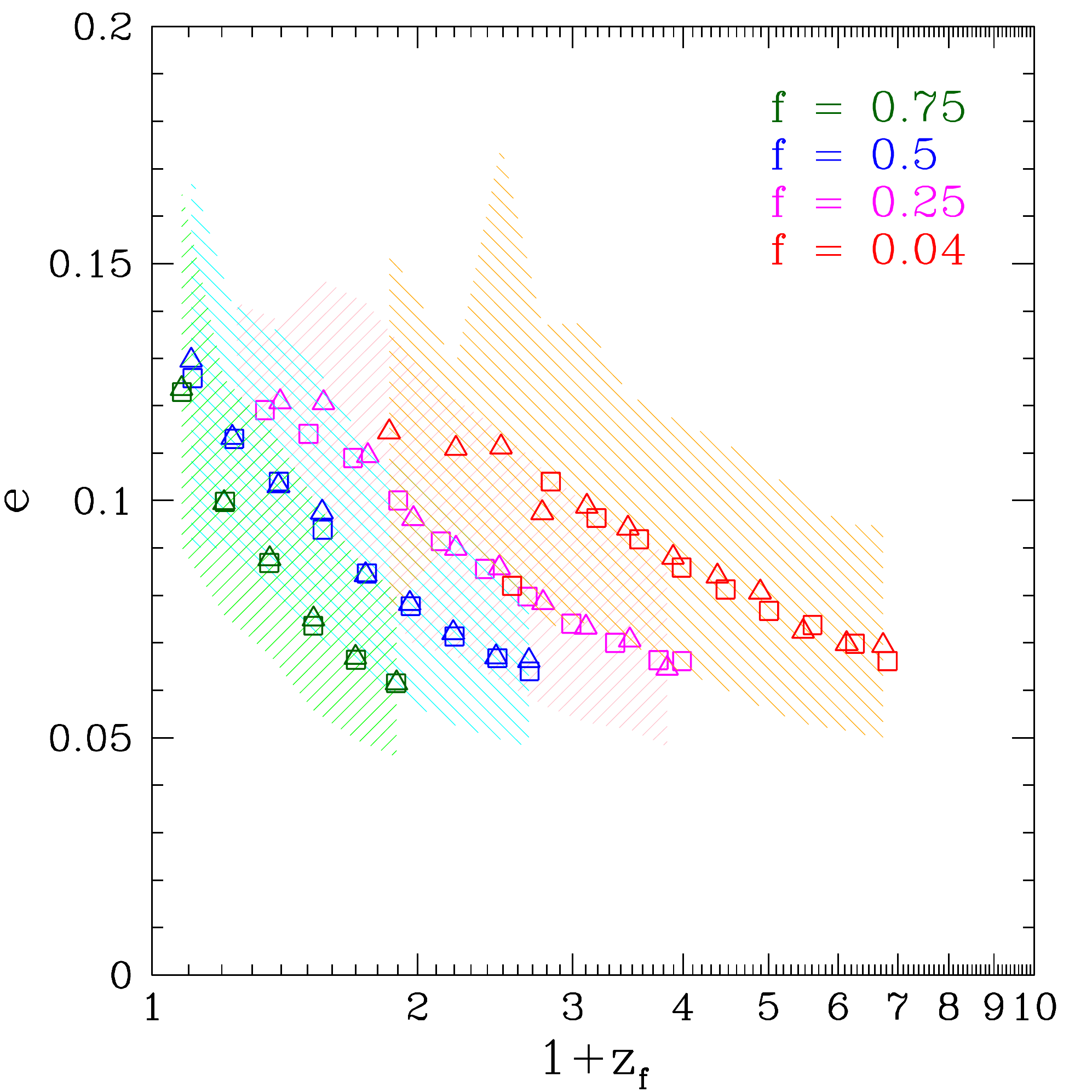}
\caption{Relation between the  ellipticity of haloes at  $z=0$ and the
  redshift  at  which  haloes  assemble different  fraction  of  their
  mass. The  points show the medians  of the distribution for  all the
  four    definition   of    $z_{f}$,   already    seen   in    Figure
  \ref{fmergertree}, while  the corresponding coloured  shaded regions
  enclose the first and third  quartile.  The squares show the results
  for  Baby,  while the  triangles  for  the  GIF2,  as in  the  other
  figures.\label{zfell}}
\end{figure}

In Figure \ref{zfell} we show  the relation between the ellipticity of
haloes at $z=0$ and their generalized formation redshifts.  The points
(squares for Baby  and triangles fo the GIF2) show  the medians of the
distribution at a  fixed $z_f$ for the four definitions,  as in Figure
\ref{fmergertree}, while the coloured shaded regions enclose the first
and third quartile.  We can see  that, using any of the definitions of
formation  redshifts, final  ellipticity  and $z_{f}$  anti-correlate:
this is  consistent with the  behaviour already seen and  discussed in
Figure \ref{cumulatives}.  It is also  interesting to notice that even
if Baby and GIF2 simulations have been run with different cosmological
parameters,  the relation  $e-z_f$ is  similar once  adopted the  same
formation redshift definition.

\section{Summary and Conclusions}\label{conclusions}

To characterise how  the distribution of halo shapes  evolves in time,
we analysed a set of three  cosmological simulations -- GIF2, Baby and
Flora, which  cover a wide range  of halo masses and  allow to compare
two  different  cosmologies.  We  presented  the  simulations and  the
post-processing  pipeline used  to  identify the  halo population  and
calculate  the shape  properties, using  the EO  halo finder,  already
described  in  \citet{despali13}.   Then we  discussed  the  resulting
distributions and proposed some best fitting relations.

The  main  result   of  this  work  is  the   existence  of  universal
distributions of the shape parameters ($e$, $p$ and the axial ratios),
when    rescaling    the    mass    to    the    universal    variable
$\nu=\delta_{c}/\sigma$.  It  allows to  eliminate the  dependences on
cosmology and epoch, moving the  distributions of all redshifts all on
the same linear relations.  Then  we report and study other properties
of halo shapes, which can be summarised as follows:
\begin{itemize}
\item  at  fixed mass,  halo  shapes  become  more elongated  at  high
  redshifts;   the  behavior   is  qualitative   the  same   for  both
  cosmologies, with  a slight difference  in the median values  due to
  the difference in  formation times of haloes;
\item  haloes   of  similar   mass  possess  larger   ellipticity  and
  prolateness  at  higher  redshifts:  on  average $e$  and  $p$  from
  redshift $z=2$ to the present time change of about $40-50\%$;
\item  at any  given  time, the  more  massive is  an  halo, the  less
  spherical it is:  this is due to the fact  that massive haloes still
  retain memory  of their  "original'' shape, which  has not  been yet
  contaminated or rounded by other events  and which is related to the
  direction of  filaments or of  the last  major merger; thus,  at any
  given time, massive haloes show higher values both of $e$ and $p$ --
  cleary the same trend is reflected in the axial ratios);
\item another  quasi-universal distribution  is given by  the relation
  between $p$ and $e$, which remains on average with a slight redshift
  dependence;
\item halo  ellipticity is  a decreasing  function of  the generalized
  formation  redshifts $z_f$  -- as  the redshift  when the  main halo
  progenitor assembles a fration $f$  of its present-day mass, with no
  particular dependence on cosmology: both GIF2 and Baby cosmology lie
  on the same relation.

\end{itemize}

To   conclude,  halo   triaxial  properties   show  a   dependence  on
cosmological   parameters  since   related   to   the  halo   assemble
histories. In this work we  have presented how ellipticity, prolatness
and axial ratios correlate with the universal variable $\nu$: in a way
that  these quantities  are  independent on  halo  mass, redshift  and
background cosmology. We find our  results useful to be implemented in
a  Monte Carlo  method to  generate  mock haloes  with given  triaxial
properties, and  in triaxial mass reconstruction  methods that require
priors for the axial ratio distributions.

\section{Acknowledgements}
We thank  Marceau Limousin,  Lauro Moscardini and  Ravi K.   Sheth for
reading  the manuscript  and  for their  useful comments.  We
  thank Marco Baldi  for his help while running  the simulations. For
this  study, GD  has  been  partially financed  by  the the  Strategic
Research   Project   $AACSE$   (Algorithms   and   Architectures   for
Computational Science  and Engineering)  of the University  of Padova.
CG's research is part of the project GLENCO, funded under the European
Seventh Framework Programme, Ideas, Grant Agreement n. 259349.

\bibliographystyle{mn2e}
%\bibliography{giulia,cgiocoli}
\bibliography{paper}
\label{lastpage}
\end{document}